\documentclass[11pt,ukrainian]{article}

\usepackage{amsmath,amssymb,enumerate}

\usepackage[mathscr]{eucal}
\usepackage{latexsym}
\usepackage{enumitem}

\usepackage{graphics}
\usepackage{hyperref}

\makeatletter \@addtoreset{equation}{section} \makeatother

\newtheorem{theorem}{Theorem}
\newtheorem{lemma}{Lemma}

\newtheorem{definition}{Definition}

\def\fracd{\displaystyle\frac}
\def\sumd{\displaystyle\sum}
\def\limd{\displaystyle\lim}

\def\prodd{\displaystyle\prod}

\begin{document}
\baselineskip 23pt \vskip 0.5 cm

\title
{EIGENVALUE DISTRIBUTION OF
LARGE WEIGHTED MULTIPARTITE RANDOM SPARSE GRAPHS}
\author{
$\;$Valentin Vengerovsky  $\!\!\:\!$, B. Verkin Institute for Low Temperature Physics and Engineering \\ of the National Academy of Sciences of Ukraine, 47 Nauky Ave., Kharkiv, 61103, Ukraine}
\date{}
\maketitle
\abstract{
We investigate the distribution of eigenvalues of weighted adjacency matrices from a specific ensemble of random graphs. We distribute $N$ vertices across a fixed number
$\kappa$ of components, with asymptotically $\alpha_j \dot N$  vertices in each component, where the vector
$(\alpha_1,\alpha_2, \ldots, \alpha_{\kappa})$ is fixed.
Consider a connected graph $\Gamma$ with $\kappa$ vertices. We construct a multipartite graph with $N$ vertices, in which all vertices in the
$i$-th component are connected to all vertices in the $j$-th component if if $\Gamma_{ij}=1$. Conversely, if $\Gamma_{ij}=0$, no edge connects the $i$-th and $j$-th components. In the resulting graph, we independently retain each edge with a probability of $p/N$, where $p$ is a fixed parameter.  To each remaining edge, we assign an independent weight with a fixed distribution, that possesses all finite moments. We establish the weak convergence in probability of a random counting measure to a non-random probability measure. Furthermore, the moments of the limiting measure can be derived from a system of recurrence relations.

}
\section{Introduction}

In recent years, interest in the spectral properties of ensembles of sparse random matrices has sharply increased. It is anticipated that the spectral characteristics of sparse random matrices will differ from those of ensembles of most matrices with independent elements (see \cite{W}, along with the survey works \cite{KKPS}, \cite{Pa:00}, and the literature cited therein).

 Significant results concerning sparse random matrices have emerged from various physical studies \cite{RB:88}, \cite{RD:90}, \cite{MF:91}, \cite{FM:96}. Notably, these works derive the equation for the Laplace transform of the limiting integrated density of states and investigate the "density-density correlator." It has been shown that there exists a critical point $p_c> 1$, around which a phase transition occurs: for $p<p_c$, all eigenvectors are localized, while for $p>p_c$, delocalized eigenvectors emerge. Unfortunately, these findings rely on replica or supersymmetry methods, necessitating a mathematically rigorous justification.

In a series of mathematical papers \cite{BG1}, \cite{BG2}, \cite{B}, it has been proven that a limit exists for the averaged moments of the integrated density of states as $N\to\infty$ in the simplest case, where the matrix elements are equal to $0$ with probability $1-p/N$ and $1$ with probability $p/N$. It is shown that the limiting moments satisfy the Carleman condition, thereby establishing the existence of a limit for the integrated density of states for the ensemble of sparse random matrices. Similar results for a broader class of sparse random matrix ensembles were obtained in \cite{KV} and \cite{KSV}. The studies in \cite{PJ} and \cite{CS} investigated delocalization and the existence of an absolutely continuous part of the limiting spectrum at zero. The behavior of the asymptotics of the correlator of moments as $ N \to\infty $ was also explored in \cite{V0}.

In \cite{V} and \cite{V1}, analogous results were obtained for the bipartite sparse random matrix ensemble. In this paper, we focus on the asymptotic behavior of the correlator of moments as $N\to\infty $ for the bipartite sparse random matrix ensemble. The rate at which these moments approach zero and the value of the leading term are crucial in physical applications. Consequently, extensive literature has been dedicated to similar studies across various ensembles of random matrices (see, for example, \cite{KKP}, \cite{APS2}, and the references therein).

In this paper, we investigate a multipartite analogue of large sparse random weighted graphs.


\section{Main results}


    We can introduce the randomly weighted
adjacency matrix of random multipartite graphs.
Let $\Xi=\{a_{ij} ,\; i \!\leq\! j,\;i,j\! \in \!{\mathbb N}\}$ be
the  set of
jointly independent identically distributed (i.i.d.) random
variables determined
  on the same probability space and possessing the moments
\begin{equation}\label{const_all_mom}
  {\mathbb E}a^{k}_{ij}\!=\!X_{k}\!<\!\infty \qquad
  \forall \; i,j,k\in {\mathbb N},
\end{equation}
where ${\mathbb E}$ denotes the mathematical expectation corresponding to
$\Xi$.
  We set $a_{ji}\!=\! a_{ij}$ for $i\!\leq\! j \;$.

Given $ 0\!<p\!\leq \!N$, let us define the family
  $D^{(p)}_N\!=\!\{d^{(N,p)}_{ij},
\; i\!\leq\! j,\; i,j\in \overline{1,N}\}$ of jointly independent
random variables
\begin{equation}
d^{(N,p)}_{ij}\!=\! \left\{ \begin{array}{ll} 1,&
\textrm{with} \ \textrm{probability } \ p/N ,
\\0,& \textrm{with} \ \textrm{probability} \ 1-p/N ,\\ \end{array}
\right.
\end{equation}
We determine $d_{ji}= d_{ij}$ and assume that $ D^{(p)}_N$
is independent from $\Xi$.

 Let us fix a natural number $\kappa\in \mathbb{N}$ and a vector $\alpha \in \mathbb{R}^\kappa$ such $\alpha_i\in (0,1), 1 \le i\le \kappa, \sumd_{i=1}^\kappa \alpha_i=1$. Let us fix a simple connected graph $\Gamma$ with $\overline{1,\kappa}$ as a set of vertices.  Let us introduce a family of nondecreasing functions  $\beta^{\overrightarrow{\alpha}}: \mathbb{N} \to \overline{1,\kappa} $, such that $\forall i\in \overline{1,\kappa} :$ $ \limd_{N\to\infty} \fracd{\# \left\{\left(\beta^{\overrightarrow{\alpha}}\right)^{-1}(i)\bigcap \overline{1,N} \right\}}{N}=\alpha_i $. Now one can consider the real symmetric $N\times N$ matrix
$A^{(N,p,\overrightarrow{\alpha})}(\omega)$:
\begin{equation}\label{dilute}
\left[A^{(N,p,\overrightarrow{\alpha})}\right]_{ij}\!=\!
\Gamma_{\beta^{\overrightarrow{\alpha}}(i),\beta^{\overrightarrow{\alpha}}(j)}\cdot a_{ij}\cdot d_{ij}^{(N,p)}
\end{equation}
that has $N$ real eigenvalues
$\lambda^{(N,p,\alpha)}_1\!\leq\!\lambda^{(N,p,\alpha)}_2 \!\leq\!\ \ldots
\!\leq\!\ \lambda^{(N,p,\alpha)}_N$.

The normalized eigenvalue counting function of  $A^{(N,p,\overrightarrow{\alpha})}$ is determined by the formula
 $$
\sigma\left({\lambda; A^{(N,p,\overrightarrow{\alpha})}}\right)\!=\!\frac{\#
\left\{j:\lambda^{(N,p,\overrightarrow{\alpha})}_j\!<\!\lambda\right\}}{N}.
$$

\begin{theorem}\label{main_thm}
Under condition
\begin{equation}\label{Karleman_cond}
X_{2m}\le \left(C \cdot m\right)^{2m}, m\in \mathbb{N}
\end{equation}
 measure $\sigma\left({\lambda; A^{(N,p,\overrightarrow{\alpha})}}\right)$ weakly converges in probability to nonrandom measure  $\sigma_{p,\overrightarrow{\alpha}}$
\begin{equation}
\sigma\left({\cdot \ ; A^{(N,p,\overrightarrow{\alpha})}}\right)\to \sigma_{p,\overrightarrow{\alpha}},\ N \to\infty ,
\end{equation}
which can be uniquely determined  by its moments
\begin{equation}\label{5}
\int \lambda^s d \sigma_{p,\overrightarrow{\alpha}}=\left\{ \begin{array}{ll}
m^{(p,\overrightarrow{\alpha})}_{k}\!=\!\sum^{\kappa}_{j=1}\sum^{t}_{i=0} S^{(j)}(k,i),&
\textrm{if } \ s=2t ,
\\0,& \textrm{if } \ s=2t-1 ,\\ \end{array} \right.
\end{equation}
where numbers  $S^{(j)}(k,i)$, can be found from the following system of recurrent relations
\begin{multline}\label{ms_r}
  \mathrm{S}^{(j)}(l,r)={\displaystyle\sum_{i=1}^{\kappa}} \Gamma_{j,i}\cdot p\cdot \sum_{f=1}^{r} {r-1\choose f-1} \cdot
   X_{2f}
         \cdot
         \\
          \sum_{u=0}^{l-r} \mathrm{S}^{(j)}(l-u-f,r-f)
       \cdot \sum_{v=0}^{u} {f+v-1\choose f-1} \cdot
       \mathrm{S}^{(i)}(u,v),
\end{multline}

with following initial conditions
\begin{equation}\label{ini_cond}
S^{(j)}(l,0)=\alpha_j\cdot \delta_{l,0}, \ j \in \overline{1,k}.
\end{equation}
\end{theorem}

The following denotations are used:
$$
{\mathcal{M}}^{(N,p,\overrightarrow{\alpha})}_t\!=\! \int \lambda^t d \sigma\left({\lambda;
A^{(N,p,\overrightarrow{\alpha})}}\right), \ M^{(N,p,\overrightarrow{\alpha})}_t=\mathbb{E}{\cal{M}}^{(N,p,\overrightarrow{\alpha})}_t,
$$
$$
C^{(N,p,\overrightarrow{\alpha})}_{t,m}= \mathbb{E}\left\{{\mathcal{M}}^{(N,p,\overrightarrow{\alpha})}_{t}\cdot {\cal{M}}^{(N,p,\overrightarrow{\alpha})}_{m} \right\}-\mathbb{E}\left\{{\cal{M}}^{(N,p,\overrightarrow{\alpha})}_{t} \right\}\cdot \mathbb{E}\left\{{\cal{M}}^{(N,p,\overrightarrow{\alpha})}_{m} \right\} .
$$
Theorem \ref{main_thm} is a corollary of Theorem \ref{thm:1}
\begin{theorem} \label{thm:1}
Assuming  conditions (\ref{Karleman_cond}),
(i) Correlators $C^{(N,p,\overrightarrow{\alpha})}_{k,m}$  vanish in the limit $\ N\to \infty$:
\begin{equation}
C^{(N,p,\overrightarrow{\alpha})}_{k,m}\le {\displaystyle\frac{C(k,m,p,\overrightarrow{\alpha})}{N}} ,  \ \forall\  k,m \in \mathbb{N}.
\end{equation}

(ii) The limit of s-th moment exists for all $s\in \mathbb{N}$:
\begin{equation}\label{5}
  \lim_{N \to \infty}M^{(N,p, \overrightarrow{\alpha})}_s =  \left\{ \begin{array}{ll}
m^{(p,\overrightarrow{\alpha})}_{k}\!=\!\sum^{\kappa}_{j=1}\sum^{t}_{i=0} S^{(j)}(k,i),&
\textrm{if} \ s=2t ,
\\0,& \textrm{if } \ s=2t-1 ,\\ \end{array} \right. ,
\end{equation}
where  numbers
  $S^{(j)}(k,i)$
are determined by  (\ref{ms_r}) -(\ref{ini_cond}).

(iii) The limiting moments $\left\{m^{(p,\overrightarrow{\alpha})}_{k}\right\}_{k=1}^{\infty}$ obey Carleman's condition
\begin{equation}
{\displaystyle\sum_{k=1}^{\infty}}{\displaystyle\frac{1}{\sqrt[2k]{m^{(p,\overrightarrow{\alpha})}_{k}}}}=\infty
\end{equation}
\end{theorem}

\section{Proof of Theorem 1}
\subsection{Walks and contributions}

\quad\ Using independence of families $\Xi$ and
$D^{(p)}_N$, we have {\setlength\arraycolsep{1pt}
\begin{eqnarray}
  M^{(N,p,\overrightarrow{\alpha})}_k\!&=&\!\int {\mathbb E} \{ \lambda^k d \sigma_{A^{(N,p,\overrightarrow{\alpha})}}
\} \!=\!{\mathbb E} \left(\frac{1}{N}\sum_{i=1}^N
[\lambda^{(N,p,\overrightarrow{\alpha})}_i]^k \right)\!=\! \frac{1}{N} {\mathbb E}\left(Tr
[A^{(N,p,\overrightarrow{\alpha})}]^k\right) \!=\! \nonumber \\ & =&\! \frac{1}{N}
\sum^{N}_{j_1=1} \sum^{N}_{j_2=1} \ldots
  \sum^{N}_{j_{k}=1} {\mathbb E} \left( A^{(N,p,\overrightarrow{\alpha})}_{j_1,j_2}
  A^{(N,p,\overrightarrow{\alpha})}_{j_2,j_3} \ldots A^{(N,p,\overrightarrow{\alpha})}_{j_{k},j_1}
  \right) \!=\! \nonumber \\
\label{base} &=&\! \frac{1}{N} \sum^{N}_{j_1=1} \sum^{N}_{j_2=1}
\ldots
  \sum^{N}_{j_{k}=1} {\mathbb E} \left( a_{j_1,j_2}
  a_{j_2,j_3} \ldots a_{j_{k},j_1} \right) \cdot \nonumber \\ & &
  \cdot {\mathbb E} \left( d^{(N,p)}_{j_1,j_2}
  d^{(N,p)}_{j_2,j_3} \ldots d^{(N,p)}_{j_{k},j_1}
  \right) \cdot  \xi^{(\overrightarrow{\alpha})}_{j_1,j_2} \cdot
  \xi^{(\overrightarrow{\alpha})}_{j_2,j_3} \cdot \ldots \cdot\xi^{(\overrightarrow{\alpha})}_{j_{k},j_1},
 \end{eqnarray}
 where
 $$
\xi^{(\overrightarrow{\alpha})}_{ij}\!=\!\Gamma_{\beta^{\overrightarrow{\alpha}}(i),\beta^{\overrightarrow{\alpha}}(j)}$$

Let $W^{(N)}_{k}$ be a set of closed walks of $k$ steps over
the set $\overline{1,N}$:
$$
W^{(N)}_{k}\!=\!\{w\!=\!(w_1,w_2,\cdots,w_k,w_{k+1}=w_1):
\forall i \!\in\! \overline{1,k+1} \;\: w_i\!\in\!
\overline{1,N}\}.
$$
For $w\!\in\!W^{(N)}_k$ let us denote
  $a(w)\!=\!\prod_{i=1}^{k} a_{w_i,w_{i+1}}$,
 $d^{(N,p)}(w)\!=\!\prod_{i=1}^{k} d^{(N,p)}_{w_i,w_{i+1}}$ and $\xi^{(\overrightarrow{\alpha})}(w)\!=\!\prod_{i=1}^{k} \xi^{(\overrightarrow{\alpha})}_{w_i,w_{i+1}}$.
Then we have
\begin{equation}\label{m_ms1}
M^{(N,p)}_k\!=\!\frac{1}{N} \sum_{w\in W^{(N)}_k} {\mathbb E} a(w)
\cdot {\mathbb E} d^{(N,p)}(w)\cdot \xi^{(\overrightarrow{\alpha})}(w).
\end{equation}
Let $w\!\in\! W^{(N)}_k$
  and
$f,g \!\in\! \overline{1,N}\ $. Denote by $n_w(f,g)$ the number
of steps $f\to g$ and $g \to f$;
$$
n_w(f,g)=\#\{i \!\in\! \overline{1,k}:\; (w_i\!=\!f \ \wedge \
w_{i+1}\! =\!g)\vee (w_i\!=\!g \ \wedge \ w_{i+1}\!=\!f)\}.
$$
Then
$$
{\mathbb E}a(w)\!=\! \prod_{f=1}^{N} \prod_{g=f}^{N} X_{n_w(f,g)}.
$$

  Given $w\!\in\! W^{(N)}_k$,
let us define the sets $V_w=\cup_{i=1}^{k}\{w_i\}$ and
$E_w=\cup_{i=1}^{k}\{(w_i,w_{i+1})\},$ where $(w_i,w_{i+1})$ is a
non-ordered pair. It is easy to see that $G_w\!=\!(V_w,E_w)$ is a
simple non-oriented graph and the walk $w$ covers the graph
$G_w$. Let us call $G_w$ the skeleton of walk $w$. We denote by
$n_w(e)$ the number of passages of the edge $e$ by the walk $w$ in
direct and inverse directions. For
$(w_j,w_{j+1})\!=\!e_j\!\in\!E_w$ let us denote
$a_{e_j}\!=\!a_{w_j,w_{j+1}}\!=\!a_{w_{j+1},w_j}$. Then we obtain
$$
  {\mathbb
E}a(w)\!=\!\prod_{e\in E_w} {\mathbb E}a^{n_w(e)}_e\!=\! \prod_{e\in
E_w} X_{n_w(e)}.
  $$
Similarly we can write
$$
  {\mathbb E}d^{(N,p)}(w)\!=\!\prod_{e\in E_w} {\mathbb
E}\left([d^{(N,p)}_e]^{n_w(e)} \right)\!=\! \prod_{e\in E_w}
\frac{p}{N}.
$$
Then, we can rewrite (\ref{m_ms1}) in the form
$$
M^{(N,p)}_k\!=\!\frac{1}{N}\sum_{w\in W^{(N)}_k} \xi^{(\overrightarrow{\alpha})}(w) \cdot \prod_{e\in E_w}
\frac{p\cdot X_{n_w(e)}}{N}\!=
$$
\begin{equation}\label{m_ms2}
=\!\sum_{w\in W^{(N)}_k} \xi^{(\overrightarrow{\alpha})}(w) \cdot\left(\frac{p^{|E_w|}}{N^{|E_w|+1}
}\prod_{e\in E_w}X_{n_w(e)} \right)\!=\!\sum_{w\in
W^{(N)}_k}\theta(w),
\end{equation}
where $\theta(w)$ is the contribution of the walk $w$ to the
mathematical
  expectation of the corresponding moment. To perform the limiting
transition
   $N\to\infty$ it is natural to separate $W^{(N)}_k$ into classes of
equivalence.
   Walks $w^{(1)}$ and $w^{(2)}$ are equivalent
  $ w^{(1)}\sim w^{(2)},\;$
if and only if there exists a bijection $\phi$ between the sets of
vertices $V_{w^{(1)}}$ and  $V_{w^{(2)}}$ such that for
$i=1,2,\ldots,k
  \;\; w^{(2)}_i\!\!=\!\!\phi(w^{(1)}_i)$
$$ w^{(1)}\sim w^{(2)}\;\Longleftrightarrow
\; \exists \phi: \ V_{w^{(1)}}\stackrel{bij}{\to}
  V_{w^{(2)}}:
  \;\forall \;i \!\in\! \overline{1,k+1}\;
\; w^{(2)}_i\!\!=\!\!\phi(w^{(1)}_i) \wedge {\beta^{\overrightarrow{\alpha}}(w^{(2)}_i)=\beta^{\overrightarrow{\alpha}}(w^{(1)}_i)}$$
  Last condition requires  every vertex and its image to be in the same component. It is essential for the further computations because the contribution of walk equals zero in the case when $\xi$ of an origin and  an end of some step equals 0. Let us denote by $[w]$ the class of equivalence of walk $w$ and by
$C^{(N)}_k$ the set of such classes. It is obvious that if two
walks $w^{(1)}$ and $w^{(2)}$ are equivalent then their
contributions are equal:
$$
  w^{(1)}\sim w^{(2)}\;\Longrightarrow
\theta(w^{(1)})\!=\!\theta(w^{(2)}).
$$
Cardinality of the class of equivalence $[w]$ is equal the number
of all mappings $\phi:V_w \to  \overline{1,N}$ such that $\forall i\in \overline{1, \kappa}$ $\phi(V_{i,w}) \subset  \left(\beta^{\overrightarrow{\alpha}}\right)^{-1}(i)$ (where $V_{i,w}=V_{w}\cap \left(\beta^{\overrightarrow{\alpha}}\right)^{-1}(i)$)  is equal to the number $\prod_{i=1}^{\kappa} \beta_{i,N}
\cdot (\beta_{i,N}-1) \cdot \ldots \cdot (\beta_{i,N}-|V_{i,w}|+1)$. Then we can rewrite
(\ref{m_ms2}) in the form
$$M^{(N,p)}_k\!=\!\sum_{w\in W^{(N)}_k}\xi^{(\overrightarrow{\alpha})}(w)
\left(\frac{p^{|E_w|}}{N^{|E_w|+1} }\prod_{e\in
E_w}X_{n_w(e)} \right)\!=$$
$$
=\!\sum_{[w]\in C^{(N)}_k} \xi^{(\overrightarrow{\alpha})}(w)\prod_{e\in E_w}X_{n_w(e)} \frac{p^{|E_w|}} {N^{|E_w|+1}
}\cdot
$$
\begin{equation}\label{m_ms3}
\left. \cdot\prod_{i=1}^{\kappa} \beta_{i,N}
\cdot (\beta_{i,N}-1) \cdot \ldots \cdot (\beta_{i,N}-|V_{i,w}|+1)\right)\!=\!\sum_{[w]\in
C^{(N)}_k} \hat{\theta}([w]).
\end{equation}
In the second line of (\ref{m_ms3}) for every class $[w]$ we choose arbitrary walk $w$ corresponding to this class of equivalence.
\subsection{Minimal and essential walks}

   Class of walks $[w]$ of $C^{(N)}_k$ has at most k vertices.
    Hence, $C^{(1)}_k
\subset C^{(2)}_k \subset \ldots \subset C^{(i)}_k \subset
\ldots C^{(k_{\beta^{\overrightarrow{\alpha}}})}_k = C^{(k_{\beta^{\overrightarrow{\alpha}}}+1)}_k= \ldots$. It is natural to denote
  $C_k=C^{(k_{\beta^{\overrightarrow{\alpha}}})}_k$. Then (\ref{m_ms3}) can be written as
\begin{equation}\label{m_ms5}
m^{(\overrightarrow{\alpha})}_k \!=\!\lim_{N \to \infty} \sum_{[w]\in C_k} \xi^{(\overrightarrow{\alpha})}(w)\cdot \prod_{i=1}^{\kappa}\alpha_i^{|V_{i,w}|}
\left(N^{|V_w|-|E_w|-1}\prod_{e\in
E_w}\fracd{ X_{n_w(e)}}{p^{-1}}\right).
\end{equation}
The set $C_k$ is finite. Regarding this and (\ref{m_ms5}), we
conclude that the class $[w]$ has non-vanishing contribution,
if $|V_w|-|E_w|-1 \!\geq \! 0$ and $w$ is a  walk  through the complete multipartite graph $K$ associated with graph $\Gamma$ and components $\{(\beta^{\overrightarrow{\alpha}})^{-1}(i)\bigcap \overline{1,N}\}_{i=1}^\kappa$. But for each simple connected
graph $G=(V,E)$  $|V_w|\! \leq \! |E_w|+1$, and the equality takes
place if and only if the graph $G$ is
  a tree.

It is convenient to deal with $\widetilde{W}^{(N,\overrightarrow{\alpha})}_{k}$ instead of $W^{(N)}_{k}$, where $\widetilde{W}^{(N,\overrightarrow{\alpha})}_{k}$
 is a set of closed walks over the set with components $\{1_1,2_1,\ldots, \#\{(\beta^{\overrightarrow{\alpha}})^{-1}(1)\bigcap \overline{1,N}\}_1\}$, $\{1_2,2_2,\ldots, \#\{(\beta^{\overrightarrow{\alpha}})^{-1}(2)\bigcap \overline{1,N}\}_2\}$ and so on.
We just renamed the vertices of the all components. Let us consider $\widetilde{C}^{(N,\overrightarrow{\alpha})}_{k}$, the set of equivalence  classes of $\widetilde{W}^{(N,\overrightarrow{\alpha})}_{k}$. As a  representative of the equivalence  class $[w]\in \widetilde{C}^{(N,\overrightarrow{\alpha})}_{k}$, we can take a minimal  walk.
\begin{definition}
   A  closed walk  $w\in \widetilde{W}^{(N,\overrightarrow{\alpha})}_{k}$ is called minimal if and only if at each stage of the passage a new vertex is the minimum element among the unused vertices of the corresponding component.
\end{definition}
Let us denote the set of all minimal walks of $\widetilde{W}^{(N,\overrightarrow{\alpha})}_{k}$ by
$\mathfrak{M}^{(N,\overrightarrow{\alpha})}_{k}$.

\noindent {\bf Example 1.} The sequence  $(1_2,1_1,1_2,1_3,2_2,2_3,1_2,2_1,3_2,1_1,1_2)$  is a   minimal walk.

\begin{definition}
The  minimal walk $w$ that has a tree as
a skeleton and has positive weight is an essential walk.
\end{definition}
Let us denote the set of all essential  walks of $\widetilde{W}^{(N)}_{k}$ by
$\mathfrak{E}^{(N)}_{k}$.
 Therefore we can rewrite
(\ref{m_ms5}) in the form
\begin{equation}\label{m_ms6}
m^{(p,\overrightarrow{\alpha})}_k \!=\! \sum_{w\in \mathfrak{E}_k} \theta(w).
\end{equation}
where
$$
\theta(w)=\prod_{j=1}^\kappa \alpha_j^{|V_jw|}
\left(\prod_{e\in E_w}\left(p\cdot  X_{n_w(e)}\right)\right).
$$

 The number of passages of each edge $e$
  belonging to the essential walk $w$ is even. Hence, the limiting
  mathematical expectation $m^{(p,\overrightarrow{\alpha})}_k$ depends only on the even
  moments of random variable  $a$. It is clear that the limiting
mathematical
  expectation $\limd_{N \to \infty}M^{(N,p,\overrightarrow{\alpha})}_{2s+1}$ is equal to zero.

\subsection{First edge splitting of essential walks}
Let us start with necessary definitions. The first vertex $w_1$
of the essential walk $w$ is called the root of the walk. We
denote it by $\rho$. Let us denote the second vertex $w_2$ of
the essential walk  $w$ by $\nu$. We denote by $l$ the half of
walk's length and by $r$ the number of steps of $w$ starting from
root $\rho$.
  In this subsection  we derive the recurrent
  relations by splitting of the walk (or of the tree) into two
  parts. To describe this procedure, it is convenient to consider
    the set of the essential walks of length $2l$ such that they
have $r$ steps starting from the root $\rho$.  We denote this set
by $\Lambda(l,r)$. By $\Lambda^{(j)}(l,r)$ we denote subset of $\Lambda(l,r)$, where the root of walk is from $j$-th component. One can see that this description is exact, in
the sense that it is  minimal and gives complete description of
the walks we need. Denote by $S^{(j)}(l,r)$   the sum of contributions of
the walk of $\Lambda^{(j)}(l,r)$ with weight $\theta$. Let us remove the edge
$(\rho,\nu)$
  from $G_w$ and denote by $\hat{G}_w$ the graph
obtained . The graph $\hat{G}_w$ has two components. Denote the
component that contains the vertex $\nu$ by $G_2$ and the
component containing the root $\rho$ by $G_1$. Add the edge
$(\rho,\nu)$ to the edge set of the tree $G_2$. Denote the result
of this operation by $\hat{G}_2$.
  Denote by $u$ the half of the walk's length
over the tree $G_2$ and by $f$ the number of steps $(\rho,\nu)$
in the walk $w$. It is clear that the following inequalities hold
for all essential walks (excepting the walk of length zero) $1\leq
f\leq r$, $r+u\leq l$. Let us denote by $\Lambda^{(j)}_1(l,r,u,f)$ the
set of the essential walks with fixed parameters $l$, $r$, $u$,
$f$ with root from j-th component. And  let us denote by $S^{(j)}_1(l,r,u,f)$  the sum of contributions of the walks
of $\Lambda^{(j)}_1(l,r,u,f)$ with weight $\theta$. Denote by $\Lambda^{(j)}_2(l,r)$ the set of
the essential walks of $\Lambda(l,r)$ such that their skeleton
has only one edge attached to the root $\rho$ from $j$-th component.
  Also we denote by $S^{(j)}_2(l,r)$ the sum of weights  $\theta$  of the
walk of $\Lambda^{(j)}_2(l,r)$. Now we can formulate the first lemma of
decomposition. It allows express $\{S^{(j)}\}_{j=1}^\kappa$ as  functions of  $\{S^{(j)}\}_{j=1}^\kappa$ and $\{S_2^{(j)}\}_{j=1}^\kappa$.

\begin{lemma}[First splitting lemma] The following relations hold for all $ j\in \overline{1,\kappa}$ \label{l1}
\begin{equation}\label{l11}
 \mathrm{S}^{(j)}(l,r)\!=\!\sum_{f=1}^{r}\sum_{u=0}^{l-r}\mathrm{S}^{(j)}_1
(l,r,u,f),
\end{equation}
where
\begin{equation}\label{l12a}
\mathrm{S}^{(j)}_1(l,r,u,f) \!=\!\alpha_j^{-1}\cdot {r-1\choose
f-1}\cdot  \mathrm{S}^{(j)}_{2}(f+u,f) \cdot
  \mathrm{S}^{(j)}(l-u-f,r-f).
\end{equation}
\end{lemma}
\

{\it Proof.} The first group of equalities is obvious. The last last group of equalities follows
from
  the bijection $F$
$$\mathrm{\Lambda}^{(j)}_1(l,r,u,f) \stackrel{bij}{\to}
\mathrm{\Lambda}^{(j)}_{2}(f+u,f) \times
  \mathrm{\Lambda}^{(j)}(l-u-f,r-f) \times $$
\begin{equation}\label{bij}
    \times \mathrm{\Theta}_{1}(r,f) ,
\end{equation}
where $\mathrm{\Theta}_1(r,f)$ is the set of sequences of 0 and 1
of
  length $r$ such that there are exactly $f$ symbols 1 in the
  sequence and the first symbol is 1.

     Let us construct this
  mapping $F$. Regarding one particular essential walk $w$ of
  $\Lambda_1(l,r,u,f)$, we consider the first edge $e_1$ of the
  graph $G_w$ and separate $w$ in two parts, the left and the
  right ones with respect to this edge $e_1$. Then we add a special
  code that determines the transitions from the left part to the
  right one and back through the root $\rho$.
  Obviously these two parts are walks, but not necessary minimal
  walks. Then  we minimize these walks. This decomposition is
  constructed by the following algorithm. We run over $w$ and
  simultaneously draw the left part, the right part, and code. If
  the current step belongs to $G_l$, we add it to the first part,
  otherwise we add this step to the second part. The code is
  constructed as follows. Each time the walk leaves the root the
  sequence is enlarged by one symbol. If
  current step is $\rho \to \nu$ this symbol is "0", otherwise  this symbol is "1".
   It is clear that
  the first element of the sequence is "1", the number of signs "1" is equal to
   $f$, and the  full length of the sequence is $r$. Now we
minimize the left and the right parts. Thus, we have constructed
the decomposition of the essential walk $w$ and the mapping $F$. The weight $\theta(w)$ of the
essential walk is  multiplicative with respect to edges and vertices. In factors $\mathrm{S}^{(j)}_{2}(f+u,f)$,
  $\mathrm{S}^{(j)}(l-u-f,r-f)$ we twice count multiplier corresponding to the root, so we need  to add factor $\alpha_j^{-1}$ in (\ref{l12a}).

\vskip 1cm

\noindent {\bf Example 2.} For
$w=(1_2,1_1,1_2,1_1,2_2,1_1,1_2,2_1,1_2,1_1,3_2,1_1,1_2,2_1,4_2,2_1,1_2,1_1,3_2,1_1,2_2,1_1,$ $2_2,1_1,1_2,2_1,1_2)$ the
left part, the right one, and the code are
$(1_2,1_1,1_2,1_1,2_2,1_1,1_2,1_1,3_2,1_1,$ $1_2,1_1,3_2,1_1,2_2,1_1,2_2,1_1,1_2)$,$\;(1_2,1_1,1_2,1_1,2_2,1_1,1_2,2_1,1_2)$,
$(1,1,0,1,0,1,0)$, respectively.

\vskip 1cm

Let us denote the left part by $( w^{(f)} )$ and the right part by
$( w^{(s)} )$. These parts are  walks with the root $\rho$.
For each edge $e$ in the tree $\hat{G}_2$ the number of passages
of $e$ of the essential walk $w$ is equal to the corresponding
number of passages of $e$ of the left part $( w^{(f)} )$. Also
for each edge $e$ belonging to the tree
  $G_1$ the number of passages of $e$
of essential walk $w$ is equal to the corresponding number of
passages of $e$ of the right part $( w^{(s)} )$. The weight of the
essential walk is  multiplicative with respect to edges. Then the
weight of the essential walk $w$ is equal to the product of
weights of left and right parts. The walk of zero length has unit
weight. Combining this with (\ref{bij}), we obtain
\begin{equation}\label{aux}
\mathrm{S}^{(j)}_1(l,r,u,f)= \alpha^{-1}\cdot\left| \mathrm{{\Theta}}_{1}(r,f) \right|
\cdot \mathrm{S}^{(j)}_2(f+u,f) \cdot \mathrm{S}^{(j)}(l-u-f,r-f),
\end{equation}

Taking into account that $|\mathrm{\Theta}_{1}(r,f)|={r-1\choose
f-1}$, we derive (\ref{l12a}) from (\ref{aux})  .

Now let us prove that for any given  elements $w^{(f)}$ of
$\mathrm{\Lambda}^{(j)}_{2}(f+u,f)$, $w^{(s)}$ of
  $\mathrm{\Lambda}^{(j)}(l-u-f,r-f)$, and the sequence $\theta \in
  \mathrm{\Theta}_{1}(r,f)$, one can construct one and only one
  element $w$ of $\mathrm{\Lambda}^{(j)}_1(l,r,u,f)$. We do this
  with the following gathering algorithm. We go along either
$w^{(f)}$ or $w^{(s)}$ and simultaneously draw the walk $w$. The
switch from $w^{(f)}$ to $w^{(s)}$ and back is governed by the
code sequence $\theta$. In fact, this procedure is inverse to the
decomposition procedure described above up to the fact that
$w^{(s)}$ is minimal. This difficulty can be easily resolved for
example by coloring vertices of $ w^{(f)} $ and $ w^{(s)} $ in
red and blue colors respectively. Certainly, the common root of $
w^{(f)} $ and $ w^{(s)} $has only one color. To illustrate
the gathering procedures we give the following example.

\vskip 1cm

\noindent {\bf Example 3.} For $ w^{(f)}
=(1_2,1_1,1_2,1_1,2_2,1_1,1_2,1_1,3_2,1_1,$ $1_2,1_1,3_2,1_1,2_2,1_1,2_2,1_1,1_2)$,$\; w^{(s)}
=(1_2,1_1,1_2,1_1,2_2,1_1,1_2,2_1,1_2), $ $\;\theta=(1,1,0,1,0,1,0)$ the gathering
procedure gives $\;w=(1_2,1_1,$ $1_2,1_1,2_2,1_1,1_2,2_1,1_2,1_1,3_2,$ $1_1,1_2,2_1,4_2,2_1,1_2,1_1,3_2,1_1,2_2,1_1,$ $2_2,1_1,1_2,2_1,1_2)$.

\vskip 1cm

It is clear that the splitting and gathering are  injective
mappings. Their domains are finite sets, and therefore the
corresponding mapping (\ref{bij}) is bijective. This completes the
proof of Lemma 1.$\blacksquare$

\vskip 1cm

To formulate Lemma \ref{l2}, let us give necessary definitions. We denote by $v$
the number of steps
  starting from the vertex $\nu$ excepting the steps
$\nu \to \rho$ and by $\Lambda^{(j)}_3(u+f,f,v)$ the
set of essential walks
   of $\Lambda^{(j)}_2(u+f,f)$ with fixed parameter $v$. Also we denote
   by $S^{(j)}_3(u+f,f,v)$  the sum of weights $\theta$ of walks of
   $\Lambda^{(j)}_3(u+f,f,v)$. Let us denote by $\Lambda^{(j,i)}_3(u+f,f,v)$ the
subset of essential walks
   of $\Lambda^{(j)}_3(u+f,f,v)$ with $\nu$ from $i$-th component. Also we denote
   by $S^{(j,i)}_3(u+f,f,v)$  the sum of weights $\theta$ of walks of
   $\Lambda^{(j,i)}_3(u+f,f,v)$. Let us denote by $G_{1,2}$ the
graph consisting of only one edge $(\rho,\nu)$, $\rho$ is from $j$-th component, $\nu$ is from $i$-th component and by
$\Lambda^{(j,i)}_4(f)$ the set of essential walks of length $2f$ such that
their skeleton coincides with the graph $G_{1,2}$. It is clear that
$\Lambda_4(f)$ consists of the only one walk $(\rho,\nu,\rho,\nu,\ldots,\nu,\rho)$
of weight $\frac{X_{2f}}{p^{-1}}$. Lemma \ref{l1} allows us to
express $\{S^{(j)}\}_{j=1}^\kappa$ as  functions of  $\{S^{(j)}\}_{j=1}^\kappa$ and $\{S_2^{(j)}\}_{j=1}^\kappa$. The next
lemma allows to express $\{S_2^{(j)}\}_{j=1}^\kappa$ as  functions of  $\{S^{(j)}\}_{j=1}^\kappa$. Thus, two lemmas allow us to express $\{S^{(j)}\}_{j=1}^\kappa$ as  functions of $\{S^{(j)}\}_{j=1}^\kappa$.
\begin{lemma}[Second splitting lemma] We have for all $j\in \overline{1,\kappa}$:\label{l2}
\begin{equation}\label{l21}
\mathrm{S}^{(j)}_2(f+u,f)=\sum_{v=0}^{u} \mathrm{S}^{(j)}_3(f+u,f,v),
\end{equation}
\begin{equation}\label{l22a}
\mathrm{S}^{(j)}_3(f+u,f,v)\!=\! \alpha_j\cdot{f+v-1\choose f-1}\cdot
\frac{X_{2f}}{p^{-1}} \cdot \sum_{i=1}^\kappa \Gamma_{ji}\cdot\mathrm{S}^{(i)}(u,v).
\end{equation}
\end{lemma}
The first group of equalities is trivial, the second group of equalities follow from the
bijection
\begin{equation}\label{bij2}
\mathrm{\Lambda}^{(j,i)}_3(f+u,f,v) \stackrel{bij}{\to}
\mathrm{\Lambda}^{(i)}(u,v)
  \times \mathrm{\Lambda}^{(j,i)}_4(f) \times \mathrm{ \Theta}_{2}(f+v,f),
\end{equation}
where
  $\mathrm{\Theta}_{2}(f+v,f)$ is the set of sequences of $0$ and $1$
of
  length $f+v$ such that there are exactly $f$ symbols 1 in the
  sequence and its last symbol  is 1. The proof is analogous to
  the proof of the first decomposition lemma. The factor $\alpha_j$ in (\ref{l22a}) is a contribution of the root in the weight.


Combining these two splitting lemmas and changing the order of
summation, we get  recurrent relations (\ref{ms_r})
with  initial conditions (\ref{ini_cond}) .

\qquad Let us denote the set of double closed walks of $k$ and $m$ steps over
 $\overline{1,N}$ by
$D^{(N)}_{k,m}\stackrel{\rm def}{\equiv}W^{(N)}_{k} \times
W^{(N)}_{m}$.
 For
$d\!=\!(w^{(1)},w^{(2)}) \!\in\!DW^{(N)}_{k,m}$ let us denote
$
a(d)\!=\!a(w^{(1)})\cdot a(w^{(2)})$,
$d^{(N,p)}(d)\!=\!d^{(N,p)}(w^{(1)})\cdot d^{(N,p)}(w^{(2)})$, $\xi^{(\overrightarrow{\alpha})}(d)\!=\!\xi^{(\overrightarrow{\alpha})}(w^{(1)})\cdot\xi^{(\overrightarrow{\alpha})}(w^{(2)})$.

 Then we obtain
$$
C^{(N,p,\overrightarrow{\alpha})}_{k,m}\!=\!\frac{1}{N^2} \sum_{d=(w^{(1)},w^{(2)})\in
D^{(N)}_{k,m}}\xi^{(\overrightarrow{\alpha})}(d) \left\{ {\mathbb E} a(d) \cdot {\mathbb E}
d^{(N,p)}(d) -\right.
$$
\begin{equation}\label{eq:cor2}
 \left.-{\mathbb E} a(w^{(1)}) \cdot {\mathbb E}
d^{(N,p)}(w^{(1)}) \cdot {\mathbb E} a(w^{(2)}) \cdot {\mathbb E}
d^{(N,p)}(w^{(2)})\right\}.
\end{equation}

For closed double walks $d\!=\!(w^{(1)},w^{(2)}) \!\in\!D^{(N)}_{k,m}$
let us denote
$
n_{d}(f,g)=n_{w^{(1)}}(f,g)+n_{w^{(2)}}(f,g).
$
Also let us introduce simple non-oriented graph $G_{d}\!=\!G_{w^{(1)}}\ \cup G_{w^{(2)}}$ for double walk $d=(w^{(1)},w^{(2)})\!\in\!D^{(N)}_{k,m}$, i.e.
$V_{d}\!=\!V_{w^{(1)}}\ \cup V_{w^{(2)}}$ and
$E_{d}\!=\!E_{w^{(1)}}\ \cup E_{w^{(2)}}$.
Then, we can rewrite \ref{eq:cor2}  in the following form
$$
C^{(N,p,\overrightarrow{\alpha})}_{k,m}\!=\! \frac{1}{N^2} \sum_{d=(w^{(1)},w^{(2)})\in
D^{(N)}_{k,m}} \xi^{(\overrightarrow{\alpha})}(d)\left\{ \prod_{e\in E_{dw}}{\mathbb E}
a_e^{n_{d}(e)} \cdot {\mathbb E} \left[d^{(N,p)}_e \right]^{n_{d}(e)} -\right.
$$
$$ -\left.\prod_{e\in
E_{w{(1)}}}{\mathbb E} a_e^{n_{w^{(1)}}(e)} \cdot {\mathbb E}
\left[ d^{(N,p)}_e\right]^{n_{w^{(1)}}(e)}  \cdot \prod_{e\in E_{w{(1)}}}{\mathbb E}
a_e^{n_{w^{(2)}}(e)} \cdot {\mathbb E} \left[ d^{(N,p)}_e\right]^{n_{w^{(2)}}(e)}\right\}\!=
$$
$$
=\! \frac{1}{N^2} \sum_{d=(w^{(1)},w^{(2)})\in
D^{(N)}_{k,m}} \xi^{(\overrightarrow{\alpha})}(dw)\left\{
\left(\frac{p}{N}\right)^{|E_{d}|}\cdot \prod_{e\in E_{d}}X_{n_{d}(e)}
 -\right.
$$
\begin{equation}\label{eq:walks}
\left. \left(\frac{p}{N}\right)^{|E_{w^{(1)}}|+|E_{w^{(2)}}|}\cdot
\prod_{e\in E_{w^{(1)}}}X_{n_{w^{(1)}}(e)}
\prod_{e\in E_{w^{(2)}}}X_{n_{w^{(2)}}(e)}\right\}.
\end{equation}

To perform the limiting
transition
   $N\to\infty$ it is natural to separate $D^{(N)}_k$ into classes of
equivalence.
   Double walks  $d=(w^{(1)},w^{(2)})$ and $u=(u^{(1)},u^{(2)})$ from  $D^{(N)}_{k,m}$ are equivalent
 if and only if their first walks  are   equivalent and their second walks  are   equivalent:

 $$
 \ d \sim u \Leftrightarrow \left(w^{(1)} \sim u^{(1)} \wedge  w^{(2)} \sim u^{(2)}\right).
 $$
 Let us denote by $[d]$ the class of equivalence of double walk $d$ and by
$\mathfrak{D}^{(N)}_k$ the set of such classes. Then we can rewrite (\ref{eq:walks})  in the following form
$$
 C^{(N,p,\overrightarrow{\alpha})}_{k,m}=\!\frac{1}{N^2} \sum_{[d]\in \mathfrak{D}^{(N)}_{k,m}}\xi^{(\overrightarrow{\alpha})}(d)
\left\{\frac{p^{|E_{d}|}}{N^{|E_{d}|}}\cdot\prod_{i=1}^{\kappa} \beta_{i,N}
\cdot (\beta_{i,N}-1) \cdot \ldots \cdot (\beta_{i,N}-|V_{i,w}|+1)\;\cdot \right.
 $$

\begin{equation}\label{eq:class}
\left.\cdot\left(
\prod_{e\in E_{d}}X_{n_{d}(e)} -\frac{p^{|E_{w^{(1)}}|+|E_{w^{(2)}}|-|E_{d}|}}
{N^{|E_{w^{(1)}}|+|E_{w^{(2)}}|-|E_{d}|}} \cdot \prod_{e\in
E_{w^{(1)}}}X_{n_{w^{(1)}}(e)} \prod_{e\in
E_{w^{(2)}}}X_{n_{w^{(2)}}(e)}\right)\right\}.
\end{equation}
Let us define a formal order of pass for double walk $d=(w^{(1)},w^{(2)})\!\in\!D^{(N)}_{k,m}$:
$$
d_i=\left\{\begin{array}{ll} w^{(1)}_i, & \textrm{if } 1 \leq
i\leq k \\ w^{(2)}_{i-k}, & \textrm{if } k+1 \leq i\leq k+m.
\end{array} \right.
$$
Let us denote the set of all minimal double walks of $D^{(N)}_{k,m}$ by $\mathfrak{M}^{(N)}_{k,m}$.
Then we obtain
  $$
 N\cdot C^{(N,p,\overrightarrow{\alpha})}_{k,m} \!=\!
\sum_{w\in \mathfrak{M}^{(N)}_{k,m}} \prod_{i=1}^{\kappa} \alpha_{i}^{|V_{i,d}|}  \lim_{N \to \infty}
\left[\frac{N^{|V_{d}|-|E_{d}|-1}}
 {p^{-|E_{d}|}}\cdot\left(
\prod_{e\in E_{d}}X_{n_{d}(e)} - \right.\right.
$$
\begin{equation}\label{eq:min_lim}
 \left. \left. \frac{p^{c(d)}} {N^{c(d)}} \cdot
\prod_{e\in E_{w^{(1)}}}X_{n_{w^{(1)}}(e)} \prod_{e\in
E_{w^{(2)}}}X_{n_{w^{(2)}}(e)}\right) \right],
\end{equation}
 where $c(d)$ is a number of common edges of $G_{w{(1)}}$ and $G_{w{(2)}}$, i.e. $c(d)=|E_{w^{(1)}}|+|E_{w^{(2)}}|-|E_{d}|$.

  $\mathfrak{M}_{k,m}$ is a finite set.  $G_{d}$ has at most 2 connected components. But if $G_{d}$ has exactly 2 connected components then  $V_{w^{(1)}}\cap V_{w^{(2)}}\!=\!\varnothing
\Rightarrow E_{w^{(1)}}\cap E_{w^{(2)}}\!=\!\varnothing
\Rightarrow c(dw)=0 \Rightarrow \left( \prod_{e\in
E_{d}}V_{n_{d}(e)} - \frac{p^{c(d)}} {N^{c(d)}} \cdot
\prod_{e\in E_{w^{(1)}}}V_{n_{w^{(1)}}(e)} \prod_{e\in
E_{w^{(2)}}}V_{n_{w^{(2)}}(e)}\right)\!=\! 0.$ Hence the contribution of such minimal double walks to $ N\cdot C^{(N,p,\overrightarrow{\alpha})}_{k,m}$ equals to 0. Otherwise  $G_{d}$ is connected graph, so
 $|V_w|-|E_w|-1\leq 0$. So (i) of  Theorem \ref{thm:1} is proved.

  Let us define a cluster  of a essential walk $w$ as a set of non-oriented edges incident to a given vertex $v\:$ of $G_w$ ($v$ is a center of the cluster). So cluster is a subgraph of the skeleton $G_w$. Also let us define an the ordered cluster of a essential walk $w$ as a cluster of the essential walk $w$ with  sequence of numbers $n_w(e_0)$, $n_w(e_1)$,\ldots, $n_w(e_l)$, where $\{e_r\}_{r=0}^{l}$ is a sequence of edges of the cluster ordered by time of their first passing.  We can derive from two splitting lemmas that number of passes covering ordered cluster with numbers $2j$, $2i_1$, $2i_2$, \ldots $2i_l$ ($s=\sum_{j=1}^{l}i_j$) equals to
  \begin{equation}\label{pass_cluster}
  n(j;i_1,i_2,\ldots,i_l)={j+s-1\choose j-1}\cdot \fracd{s!\prodd_{r=1}^{l}i_r}{s\cdot(s-i_1)\cdot(s-i_1-i_2)\cdot \ldots \cdot i_l\cdot \prodd_{r=1}^{l}i_r!}.
  \end{equation}
  Indeed, steps to  center of ordered cluster $v$ uniquely determined by choice of steps from the vertex $v$. We can choose steps from $v$ along edge $e_0$ by ${j+s-1\choose j-1}$ ways. After that we can choose steps from $v$ along edge $e_1$ by ${s-1\choose i_1-1}$ ways. After that we can choose steps from $v$ along edge $e_2$ by ${s-i_1-1\choose i_2-1}$ ways and so on. Thus the number of passes covering ordered cluster with numbers $2j$, $2i_1$, $2i_2$, \ldots $2i_l$ ($s=\sum_{j=1}^{l}i_j$)  is equal to
   \begin{equation}\label{pass_cluster1}
  {j+s-1\choose j-1}\cdot {s-1\choose i_1-1}\cdot {s-i_1-1\choose i_2-1}\cdot \ldots \cdot{s-i_1-i_2-\ldots -i_{l-1}-1\choose i_l-1}={j+s-1\choose j-1}\cdot
  \end{equation}
  $$
   \cdot \fracd{s!\cdot i_1}{s\cdot(s-i_1)!\cdot i_1!}\cdot \fracd{(s-i_1)!\cdot i_2}{(s-i_1)\cdot(s-i_1-i_2)!\cdot i_2!}\cdot \ldots\cdot\fracd{(s-i_1-i_2-\ldots-i_{l-1})!\cdot i_l}{(s-i_1-i_2-\ldots-i_{l-1})\cdot 0!\cdot i_l!}
  $$
  But it is evident that numbers in (\ref{pass_cluster}) and (\ref{pass_cluster1}) coincide.

  Let us define the weight of the ordered cluster by
 \begin{equation}\label{weight_cluster}
 \theta(j;i_1,i_2,\ldots,i_l)=n(j;i_1,i_2,\ldots,i_l)\cdot\prodd_{r=1}^{l} (p\cdot X_{2i_r}).
  \end{equation}
  Combining (\ref{weight_cluster}) with (\ref{Karleman_cond}) and estimate
  $C_1 k^ke^{-k}\le k!\le C_2 k^ke^{-k}$, we obtain the following estimate
  \begin{equation}\label{est_weight_cluster}
 \theta(j;i_1,i_2,\ldots,i_l)\le 2^j C_3^ss^{2s}(1+p)^s.
  \end{equation}
  Let us define an ordered skeleton as a skeleton with all ordered clusters and chosen root. Let us define the weight of  an ordered skeleton as the number of passes covering it multiplied by $\prodd_{e\in E_w}(p\cdot X_{n_w(e)})$. From (\ref{est_weight_cluster}) we get that the weight of ordered skeleton of essential walk $w$ from $\mathfrak{E}_k$ is not greater than $C_4^kk^{2k}(1+p)^k$. But the number of all ordered skeleton is not greater than $\kappa^{k+1}\sumd_{i=0}^k\left(\fracd{2i!}{i!(i+1)!}\cdot{k-1\choose i-1}\right)\le \kappa^{k+1}\cdot 2^{3k}$. So (iii) of Theorem \ref{thm:1} is proved.


\begin{thebibliography}{99}




\bibitem{W} E.P.Wigner. On the distribution of the roots of
certain symmetric matrices, Ann.Math. {\bf 67}: (1958), 325-327.




\bibitem{KKPS}  A. Khorunzhy, B. Khoruzhenko, L. Pastur and
M. Shcherbina. The Large-n Limit in Statistical Mechanics and
Spectral Theory
     of Disordered Systems.
     Phase transition and critical phenomena.v.15, p.73, Academic
     Press, 1992

\bibitem{Pa:00} Pastur, L.: Random matrices as paradigm. In: Fokas,
A.,
Grigoryan, A., Kibble, T., Zegarlinski B. (eds.)
\emph{Mathematical Physics 2000}. London: Imperial College Press,
(2000), pp. 216-266.

\bibitem{RB:88} G.J. Rodgers  and A.J. Bray. Density of states of a
sparse random matrix, Phys.Rev.B {\bf 37}, (1988), 3557-3562.

\bibitem{RD:90}  G.J. Rodgers and C. De Dominicis. Density of states of
sparse random matrices,
J.Phys.A:Math.Jen.{\bf 23}, (1990), 1567-1566.

\bibitem{MF:91} A.D.Mirlin, Y.V.Fyodorov. Universality of the
level correlation function of sparce random matrices,
  J.Phys.A:Math.Jen.{\bf 24}, (1991), 2273-2286.

\bibitem{FM:96} Y.V.Fyodorov, A.D.Mirlin. Strong eigenfunction
correlations near the Anderson localization transition.
arXiv:cond-mat/9612218 v1


\bibitem{BG1} M.Bauer and O.Golinelli. Random incidence matrices:
spectral density at zero energy, Saclay preprint T00/087;
cond-mat/0006472

\bibitem{BG2} M.Bauer and O.Golinelli. Random incidedence matrices:
moments and
spectral density, J.Stat. Phys. {\bf 103}, 301-336, 2001


\bibitem{B} B. Bollobas {\it Random Graphs }   Acad. Press (1985)




























\bibitem{KV}  A. Khorunzhy, V. Vengerovsky. On asymptotic solvability
of random graph's laplacians. Preprint \textit{lanl.arXiv.org}
math-ph/0009028

\bibitem{KSV} Khorunzhy O., Shcherbina M., and Vengerovsky V. Eigenvalue distribution of large weighted random graphs, J. Math. Phys. {\bf 45}  N.4: (2004), 1648-1672.



\bibitem{Me:91} M.L.Mehta: \emph{Random Matrices}. New York: Academic
Press, 1991







\bibitem{V0} V. Vengerovsky. Asymptotics of the correlator of some ensemble of sparse random matrices,
JMPAG {\bf 04-2}, 2016, 135-160.


\bibitem{V} V. Vengerovsky. Eigenvalue Distribution of a Large Weighted Bipartite Random Graph,
JMPAG {\bf 10-2}, 2014, 240-255.

\bibitem{V1} V. Vengerovsky. Eigenvalue Distribution of a Large Weighted Bipartite Random Graph. Resolvent Approach,
JMPAG {\bf 12-1}, 2016, 78-93.



\bibitem{KKP} { A. Khorunzhy, B. Khoruzhenko, and L. Pastur},
{\textnormal Asymptotic properties of large random matrices with independent
entries}, { J. Math. Phys.}{\textnormal(1996), v. 37, p.
5033-5059.}

\bibitem{APS2} { S. Albeverio, L. Pastur, M. Shcherbina},{\textnormal On the $1/n$
expansion for some unitary invariant ensembles of random
matrices},
{ Commun. Math. Phys.}{\textnormal (2001), v. 224,  p. 271-305.}

\bibitem{PJ} { Paul Jung, Jaehun Lee},
{\textnormal Delocalization and limiting spectral distribution of
Erdos-Renyi graphs with constant expected degree}, { Electron. Commun. Probab. }  {\textnormal (2018), v. 23,  p. 1-13.}

\bibitem{CS} { Simon Coste, Justin Salez},
{\textnormal Emergence of extended states at zero in the spectrum of sparse random graphs}, { The Annals of Probability }  {\textnormal (2021), v.  49 (4),  p. 2012 - 2030.}
\end{thebibliography}
\end{document}